\documentclass[12pt,preprint]{aastex}
\begin{document}

\newcommand{\kms}{$\,\mbox{km}\,\mbox{s}^{-1}$}
\newcommand{\etal}{et al.}
\newcommand{\HI}{H{\sc i}\ }
\newcommand{\LCDM}{$\Lambda$CDM}
\newcommand{\MLs}{\ensuremath{\Upsilon_{\star}}}
\newcommand{\Om}{\ensuremath{\Omega_m}}
\newcommand{\OL}{\ensuremath{\Omega_{\Lambda}}}
\newcommand{\norm}{\ensuremath{\sigma_8}}

%\shorttitile{Cosmological Limits from LSB Galaxies}
%\shortauthors{McGaugh, Barker, \& de Blok}

\title{A Limit on the Cosmological Mass Density and Power Spectrum
from the Rotation Curves of Low Surface Brightness Galaxies}

\author{Stacy S.~McGaugh\altaffilmark{1}}

\author{Michael K.~Barker\altaffilmark{1,2}}

\and

\author{W.J.G.~de~Blok\altaffilmark{3,4}}

\altaffiltext{1}{Department of Astronomy, University of Maryland,
	College Park, MD 20742-2421}
%\email{ssm@astro.umd.edu}
\altaffiltext{2}{Department of Astronomy, University of Florida,
	Gainesville, FL 32611-205}
%\email{mbarker@astro.ufl.edu}
\altaffiltext{3}{Australia Telescope National Facility,
	PO Box 76, Epping NSW 1710, Australia}
%\email{edeblok@atnf.csiro.au}
\altaffiltext{4}{Bolton Fellow}

\begin{abstract} 
The concentrations of the cuspy dark matter halos predicted by simulations
of cold dark matter are related to the cosmology in which the halos form.
Observational constraints on halo concentration therefore map into
constraints on cosmological parameters.  In order to explain the observed
concentrations of dark matter dominated low surface brightness galaxies,
we require a cosmology with rather little power on galaxy scales. 
Formally, we require $\sigma_8 \Gamma_{0.6} < 0.23$,
where $\Gamma_{0.6}$ is a modified shape parameter appropriate to this problem. 
Practically, this means either $\Omega_m < 0.2$ or $\sigma_8 < 0.8$.
These limits apply as long as we insist that the cuspy halos found in
simulations must describe the halos of low surface brightness galaxies.
A low density cosmology helps with the low observed concentrations,
but it offers no explanation of the many cases where the shape of the density
profile deviates from the predicted cuspy form.
These cases must have suffered very extensive mass redistribution if the
current halo formation picture is not to fail outright.  It is far from
clear whether any of the mass redistribution mechanisms which
have been suggested (e.g., feedback) are viable.
\end{abstract}


\section{Introduction}

There has been considerable recent interest in the inner structure of
dark matter halos.  Simulations of structure formation with cold
dark matter (CDM) have advanced to the point where they are
able to predict the density profile of dark matter halos (e.g., Navarro,
Frenk, \& White 1997; hereafter NFW).  The predicted form of these
NFW halos is fairly universal, with the central halo concentrations
depending upon the density of the universe at the time of
halo formation.  This depends on the cosmology, so a constraint on halo
concentrations translates into one on cosmological parameters. 

The rotation curves of spiral galaxies provide accurate tracers of
the combined potential of dark and luminous mass.  As such, they
should provide a measure of concentration which can be mapped to
cosmology.  Unfortunately, the simulations provide accurate predictions
only for the dark matter.  Luminous mass can be a significant fraction of the
total mass at small radii in bright spirals, making it difficult to
extract a measure of the dark-only mass which simulations predict.

Fortunately, there is a class of dark matter dominated, low surface brightness
(LSB) galaxies ideally suited for this experiment.  The luminous mass of
these objects is very diffuse, greatly reducing the impact of the stellar
mass on the inferred halo parameters (e.g., de Blok \& McGaugh 1997).
In these systems, the luminous mass merely provides a 
convenient tracer of the dominant dark mass. 

Useful constraints on the dark matter distribution require accurate,
high resolution data.  McGaugh, Rubin, \& de Blok
(2001) present a large sample of rotation curves for LSB galaxies with
resolution sufficient to constrain the NFW concentration parameter.
We have modeled these in detail (de Blok, McGaugh, \& Rubin 2001;
hereafter BMR).  Further such data have been obtained by de Blok \&
Bosma (2002; hereafter BB).  In this paper,
we use the results of BMR and BB to obtain a constraint on cosmology.

BMR and BB find that cuspy NFW halos provide a poor description of the data. 
The most obvious interpretation is that real dark matter halos
do not have the cusps.
Halo models with constant density cores are more viable.

Nonetheless, cuspy halos do seem to be the generic prediction
of CDM simulations, whereas soft-core isothermal halos enjoy no such theoretical
basis or cosmological context.  Moreover, it is often possible to find NFW fits
to the rotation curves of individual galaxies, even if they are formally worse
than isothermal halo fits.  There is therefore
some reason to persist with NFW halos despite the substantial observational
evidence against them.  The purpose of this
paper is to illustrate the consequences for cosmology if we insist
that dark matter halos must have the cuspy NFW form.

In \S 2 we review the data, its interpretation, and
the applicability of the NFW form.
In \S 3 we discuss the relation of CDM halos 
to cosmological parameters and combine this with observations to
place limits on cosmology.  The implications of these limits are
discussed in \S 4, as are some possible ways out. 
Conclusions are given in \S 5.

\section{Data}

The data we use here are the NFW fits to the high resolution hybrid
H$\alpha$-\HI\ rotation curves of LSB galaxies described by BMR and BB. 
These are based on recent optical data (McGaugh et al.\ 2001; BB)
combined with \HI\ data (van der Hulst \etal\ 1993; de Blok,
McGaugh, \& van der Hulst 1996; Swaters et al.\ 2002a)
in order to optimize both the
resolution and the radial extent of the data (see BMR).  The observed rate
of rise of these rotation curves is rather gradual compared to the rate of
rise predicted by NFW halos for a broad range of plausible cosmologies
(Fig.~1). 

\begin{figure}
\plotone{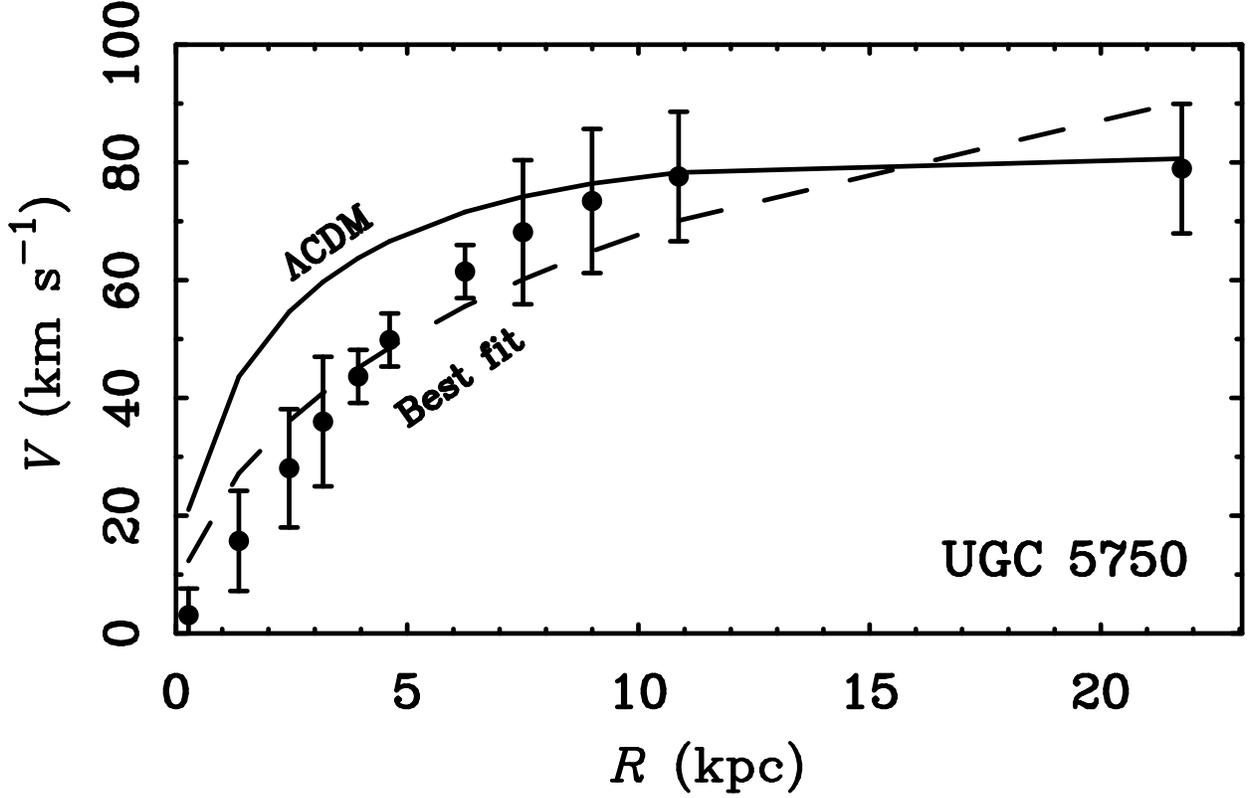}
\caption{The rotation curve of the low surface brightness galaxy UGC 5750.
Also shown are the best fitting NFW halo parameters
($c=2.6$, $V_{200} = 123$ km s$^{-1}$:  dashed line)
for the limiting case of a zero mass (minimum) disk,
and what the NFW halo should look like for a galaxy of this rotation
velocity in the standard $\Lambda$CDM cosmology
($c=10$, $V_{200} = 67$ km s$^{-1}$:  solid line).
The excess of the solid line over the data illustrates the
cuspy halo problem. Though an NFW fit can be made (dashed line),
it is a poor description of the data, and requires a very
low concentration ($c=2.6$ does not occur in any plausible cosmology).
These problems become more severe as allowance is made for stars (BMR; BB).
\label{f1}}
\end{figure}

\placefigure{f1}

It was noted some time ago that the solid body nature of the rotation
curves of low luminosity galaxies posed serious difficulties for cuspy
halos (Moore 1994; Flores \& Primack 1994; McGaugh \& de Blok 1998a).
A possible explanation offered for this was that the 21 cm data
were adversely affected by beam smearing (van den Bosch et al.\ 2000). 
This might diffuse away the sharp signature of
a cuspy halo, but is no guarantee that a cusp is present (van den Bosch
\& Swaters 2001).

The issue of beam smearing can be addressed by improving
the resolution of the observations.  McGaugh et al.\ (2001) and BB
obtained H$\alpha$ long slit data with 1'' - 2'' spatial resolution. 
This improves on the previous 21 cm data by more than an order of magnitude.
For the most part, the high resolution optical data confirms the 
gradual rate of rise of the rotation curves of LSB galaxies which had
previously been indicated by the 21 cm data.  
Reports of the severity of beam smearing have been greatly exaggerated.  

This is not to say that beam smearing was insignificant in all cases;
there are certainly a few where it is apparent.  However, it was not
pervasive enough to hide the very distinctive signature of the cusps
which are predicted for halos in the standard \LCDM\ cosmology.
Moreover, the objections of McGaugh \& de Blok (1998a,b) to the CDM
explanation for rotation curves were not based on detailed NFW
fits to 21 cm data, but rather on the systematics of the shapes of
rotation curves.  These are well predicted by the light distribution, even
in LSB galaxies where the luminous mass is insignificant.  The cases
where beam smearing was significant (e.g., F574-1: Swaters, Madore, \&
Trewhella 2000) had been discrepant from these relations; the improved
optical data remedy this.  This makes the problems with CDM discussed by
McGaugh \& de Blok (1998a) more severe, not less.

BMR and BB show that flat core, isothermal halos always provide a better
description of the data than do NFW halos. 
There is a remarkable unanimity of results on this point.
All high resolution rotation curve data show a strong preference for
soft cores (Marchesini et al.\ 2002; Bolatto et al.\ 2002;
Blais-Oullette et al.\ 2001; Salucci 2001;
Borriello \& Salucci 2001; C\^ot\'e et al.\ 2000).
So far as we are aware, there is no contradictory observational evidence
which might prefer cuspy halos in galaxies which are dark matter dominated. 

There are claims (e.g., Jimenez, Verde, \& Oh 2002) that cuspy
halos might do as well as halos with soft cores in high surface
brightness (HSB) galaxies, but these are not obviously dark matter dominated
(e.g., Palunas \& Williams 2000).  The role of stellar mass
is a great impediment to any conclusions about cuspy halos in HSB galaxies.
In cases where constraints beyond the rotation curve can be placed on
stellar mass, it is very difficult to fit cuspy halos 
(e.g., Weiner, Sellwood, \& Williams 2001).
This is especially true in the Milky Way where many lines of evidence
exclude a cuspy halo (Binney \& Evans 2001;  Bissantz \& Gerhard 2001). 
For LSB galaxies, Jimenez \etal\ (2002) confirm the results of BMR. 

The preference of the data for soft cores in LSB galaxies can no longer
be attributed
to resolution effects as suggested by van den Bosch et al.\ (2000). 
There are now many well resolved rotation curves
for dark matter dominated galaxies from a number of independent sources.
Contrary to the expectations of beam-smearing arguments, the best resolved
rotation curves prove to be the least consistent with
cuspy halos (de Blok et al.\ 2001).  It is unlikely that any significant
systematic effect afflicts the optical data as the rotation curves obtained
for the same galaxies by independent observers are in good agreement
(McGaugh \etal\ 2001; BB). 

Given the strong observational evidence against cuspy halos, it is perhaps
puzzling that their use persists.  This is largely attributable to the
convergence of many theoretical results: cuspy halos do appear to be a
fundamental prediction of CDM (see below).  Moreover,
while NFW fits to rotation curve data are distinctly inferior to fits with
isothermal halos, it is usually possible to find some combination
of NFW parameters which more or less go through the data (Fig.~1). 
Given the importance of cuspy halos to the prevailing paradigm,
and the lack of theoretical basis for halos with soft cores,
the preference of the data for the latter might not, in itself,
be seen as completely disastrous.

Fortunately, it is possible to apply further tests.
In CDM models the parameters of NFW halos have a direct relation to cosmology.
Not only must the NFW form fit (however crudely) individual rotation curves,
but the statistics of the parameters of these fits must be consistent with
the cosmology which gave rise to them.  The work of BMR and BB provides, for
the first time, a large, homogeneous sample of
precise, high spatial resolution rotation curve data for dark matter dominated
LSB galaxies to which this test can be applied.
Here, we follow through on this to see what these data imply for cosmology.

\section{Halo Parameters and Cosmology}

\subsection{Theoretical Framework: NFW Halos}

Cosmological simulations make clear predictions about the
radial density distribution of CDM halos (e.g., NFW).  These have a
simple form which has come to be known as the NFW halo.  An NFW
halo is specified by two parameters:  a concentration $c$ and a scale such
as the circular velocity at the virial radius,
$V_{200}$.  The rotation curve due to an NFW halo is
\begin{equation}
V_c^2(x)=V_{200}^2 \frac{\ln(1+cx)-cx(1+cx)^{-1}}
{x[\ln(1+c)-c(1+c)^{-1}]},
\end{equation} 
where $x=R/R_{200}$ and $R_{200}$ (in kpc) $= V_{200}
h^{-1}$ (in \kms).  Fits of this form to rotation curves are very sensitive
to the concentration parameter $c$, which itself depends on cosmological
parameters. 

The precise form of the inner density profile is critical to the result, so this
warrants further examination.  The profiles of halos predicted by CDM
have now been studied by many groups (Dubinski 1994;
Cole \& Lacey 1996; NFW; Tormen, Bouchet, \& White 1997; Moore et al.\
1998, 1999; Tissera \& Dominguez-Tenreiro 1998; Nusser \& Sheth
1998; Syer \& White 1998; Avila-Reese,
Firmiani, \& Hernandez 1998; Salvador-\'Sole, Solanes,
\& Manrique 1998; Jing 2000; Jing \& Suto 2000; Kull 1999).   
These investigations all either concur with the NFW halo profile,
or find an even steeper cusp in the inner parts.  If we parameterize the
inner slope of the density profile by $\gamma$ so that
$\rho \propto r^{-\gamma}$, then the NFW result is $\gamma = 1$.  
Moore \etal\ (1998, 1999) find a steeper inner slope ($\gamma = 1.5$)
from high resolution simulations.  It is unclear whether this difference in
$\gamma$ is significant, or more a matter of the choice of fitting
function.  If the inner slope is very steep ($\gamma > 1$), more severe
limits would result than those we derive here for the NFW case.  Given
this, and the clear mapping between NFW halo parameters and
cosmological ones, we restrict the analysis to the NFW case ($\gamma = 1$)
as the conservative limit.

There are some differences between these theoretical studies,
but from an observational perspective these are unimportant.  The one real
exception to the many studies predicting $\gamma \ge 1$ is the study of
Kravtsov et al.\ (1998).  They find\footnote{In addition to their analysis
of their simulations, Kravtsov et al.\ (1998) examined the \HI\ data
available at the time, and found that it indicated $\gamma \approx 0.2$.
This same value is found in subsequent optical data (de Blok et al.\ 2001).
The agreement between these results is another indication that systematic
errors are not significant.} a shallower inner
halo slope than NFW, with $\gamma \approx 0.4$.  This value is
significantly discrepant from other studies, and was estimated perilously
close to the limits of the simulations.
These workers have now retracted the claim of shallow inner halo profiles
(Klypin et al.\ 2001), and concur with other studies that the predicted
slope is $\gamma \ge 1$. 

If we treat $\gamma$ as a free parameter, the data give a median
$\gamma \approx 0.2$, with essentially all well resolved data being
consistent with $\gamma = 0$ (de Blok \etal\ 2001).  Though it is
possible to fit NFW halos to much of the data, it is not the form
favored by the data (\S 2).
If CDM simulations produced a halo with a soft core, there would
be little implication for cosmology.  However, the vast majority of
theoretical work on this subject predicts $\gamma \ge 1$, with no room to
treat $\gamma$ as a free parameter.  If we accept this as the correct
prediction of CDM, then the observed distribution of concentrations
places interesting limits on cosmology.

\subsection{The Halo Concentration--Cosmology Connection}

The characteristic concentration of CDM halos depends on the density of the
universe at the time of halo formation (NFW).  This in turn depends on the
density parameter (\Om), the distance scale ($h$), and the amplitude of
the power spectrum on the relevant scales.  The latter depends on both
the normalization (\norm) and shape of the power spectrum.

NFW provide a prescription which relates halo parameters to the cosmology in
which they form.  This is encapsulated by the program {CHARDEN}
provided by Navarro (private communication).  For a specified halo mass
and set of cosmological parameters, {CHARDEN} gives the
concentration and other parameters.  We use {CHARDEN} to make
several hundred realizations of the mean concentration parameter for a
variety of cosmologies.  The grid of realizations samples
parameters in the ranges $0.1 < \Omega_m < 0.5$,
$0.45 < h < 0.85$, $0.2 < \sigma_8 < 2$, and $0.75 < n < 1.25$,
where $n$ is the scalar spectral index of the power spectrum.
We consider only flat ($\Om + \OL = 1$) cosmologies. 
Open cosmologies give somewhat higher
concentrations, all other things being equal.  Since LSB galaxies
require rather low concentrations, the constraints on an open cosmology
would be tighter.

The results from {CHARDEN} are shown in Fig.\ 2, which plots the
concentration parameter $c$ as a function of cosmic parameters.
The concentration also depends weakly on halo mass, which could be plotted
as a third axis.  We show the case for $V_{200} = 163$ \kms\ 
($M = 10^{12} M_{\odot}$).  This mass scale is appropriate to bright galaxies,
and is somewhat higher than is typical for the LSB galaxies in our
sample.  The correlation of $c$ with $V_{200}$ goes in the sense that
$c$ is higher for lower $V_{200}$, so adopting this mass scale for
mapping between theory and observation is a conservative choice. 
That is, a lower mass halo more appropriate for an LSB galaxy would
have a higher concentration.  When required to match the low observed
concentrations, a more stringent constraint on cosmology would be implied.

\begin{figure}
\epsscale{0.6}
\plotone{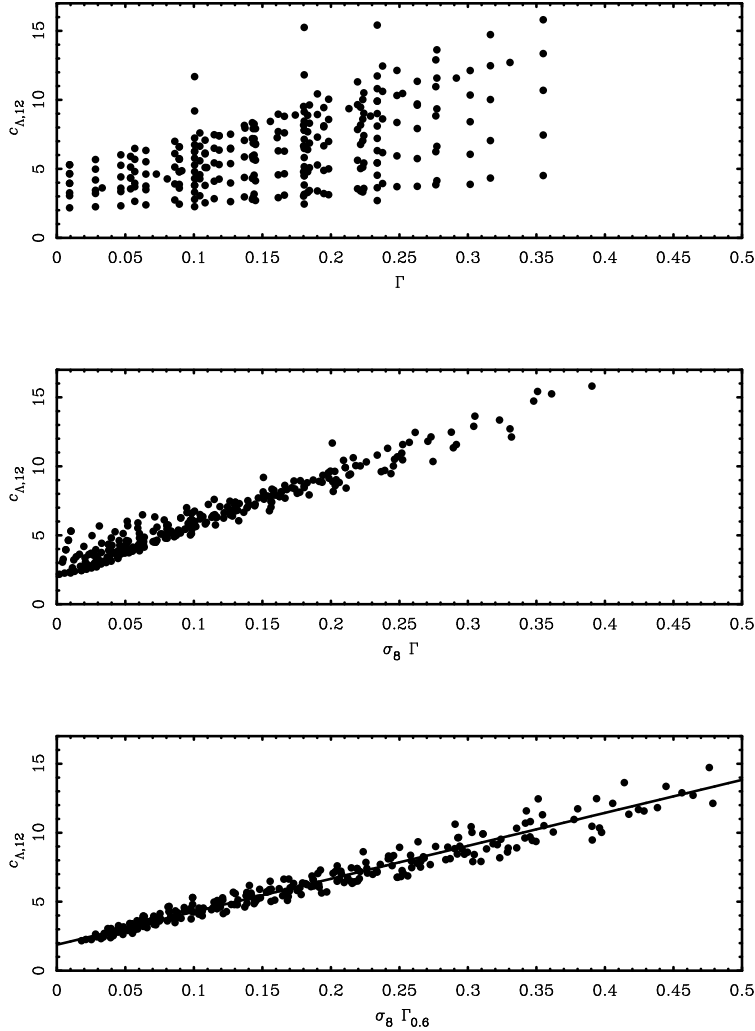}
\caption{
The concentration parameter $c_{\Lambda,12}$ predicted for a $10^{12} M_{\sun}$
NFW halo in a flat universe as a function of cosmic parameters.
In the top panel we show the concentration as a function of the shape
parameter $\Gamma$, which depends on $\Omega_m$, $h$, $\Omega_b$, and $n$.
There is a clear correlation but the realizations are stirated
by their different normalizations.  This is remedied by considering
the product of the shape parameter and the normalization, $\sigma_8$
(middle panel).  There is still some residual scatter which is reduced by use
of the modified shape parameter, $\Gamma_{0.6}$ (bottom panel). 
This depends on $\Omega_m^{0.6}$
rather than the linear power of $\Omega_m$ which appears in the standard
shape parameter (equations 2 and 3).  The line
in the bottom panel (equation 4) is a fit to these realizations
which we use to relate the observed concentrations to cosmology.
\label{f2}}
\end{figure}

\placefigure{f2}

The top panel of Fig.\ 2 shows the dependence of $c$ on the shape
parameter $\Gamma$ which to a first approximation is the product
$\Om h$.  Here we use the common fitting formula 
\begin{equation}
\Gamma = \Om h \; e^{-\left(\Omega_b+\sqrt{2h} \frac{\Omega_b}{\Om}\right)}
-0.32\left(n^{-1}-1\right)
\end{equation}
(e.g., White \etal\ 1996).
The baryon density $\Omega_b$ acts as small correction
factor to the shape of the power spectrum for a given $\Om h$. 
An increase in $\Omega_b$
depresses the relative amount of power on smaller
scales, thus acting to lower halo concentrations.
A similar effect can be obtained by tilting the spectrum ($n < 1$),
but these are rather weak effects compared to $\Om h$.

There is a clear call for a second parameter in the top panel of Fig.~2.
This is the normalization of the power spectrum \norm\ (middle panel). 
The combination $\norm \Gamma$ seems to account for most components
of the dependence of halo concentration on cosmology.  There does remain
some stiration at low concentration and more scatter than one would like.
Motivated by the fact that the relevant dependence on the mass density in
many structure formation problems is $\Om^{0.6}$, we define a modified
shape parameter $\Gamma_{0.6}$:
\begin{equation}
\Gamma_{0.6} = \Om^{0.6} h\; e^{-\left(\Omega_b+\sqrt{2h}
\frac{\Omega_b}{\Om}\right)} -0.32\left(n^{-1}-1\right).
\end{equation}
This is identical to the usual shape parameter with the exception
of the power of \Om.  The result
is a linear dependence of halo concentration on the product
$\norm \Gamma_{0.6}$ (bottom panel of Fig.~2).

The realizations in the bottom panel of Fig.\ 2 are well described by
the relation 
\begin{equation}
c_{\Lambda,12} = 1.88+23.9 \sigma_8 \Gamma_{0.6}
\end{equation}
where $c_{\Lambda,12}$ is specific to
$10^{12} M_{\odot}$ NFW halos in flat $\Lambda$CDM cosmologies.  The
residual scatter about this relation is small, with a standard deviation
of 0.084 in the ratio of realized to predicted concentration.
This relation for the characteristic
concentration produced by a given cosmology can be used to map
the observed concentrations to a limit on cosmology (\S 3.4).

\subsection{Scatter in Concentrations}

Equation (4) gives the characteristic concentration of NFW halos
for a given set of cosmological parameters.  Of course, not all halos of
a given mass are identical ---
some scatter is expected about the nominal concentration. 
This issue has been
investigated in detail by Jing (2000) and Bullock et al. (2001).  Both
groups find a lognormal distribution of concentration parameters
with $\sigma_c = 0.18$.  This apparent consistency is
marred by the fact that Jing (2000) uses the natural logarithm and Bullock
et al.\ (2001) use the base 10 logarithm so that the same value of
$\sigma_c$ actually corresponds to a broader distribution in the
latter case.  Nevertheless, there does seem to be broad agreement about
the form the distribution should take even if the value of $\sigma_c$ is
not uniquely fixed by simulations.

The histogram of observed concentrations is shown in Fig.\ 3.
These data are taken from BMR and BB for the case of minimum disk.
This choice is made to maximize the number of galaxies which can be
included in the analysis, and to be conservative in the sense that minimum disk
fits allow larger concentrations than do fits which include stars.
There are other sources of data which we do not include here in order to
maintain a consistent analysis:
the mass models of BMR and BB have been constructed 
in an identical fashion.  Other data for dark matter dominated galaxies
are limited in number so far, so their inclusion or exclusion makes little
difference to the histogram in Fig.~3.  As discussed in \S 2, the independent
analyses which do exist are entirely consistent with the findings of
BMR and BB.

The distribution in Fig.\ 3 is broad: there are many galaxies with low ($c<6$)
concentrations, and some with rather high concentrations ($c > 15$).
Though these are outside the nominal range one would expect for \LCDM,
there is a well defined central peak.  There is, therefore, some hope that
the anticipated lognormal distribution can be fit to the data if the
relevant cosmological parameters are allowed to vary, and an explanation
can be found for the outlyers.

\begin{figure}
\epsscale{1.0}
\plotone{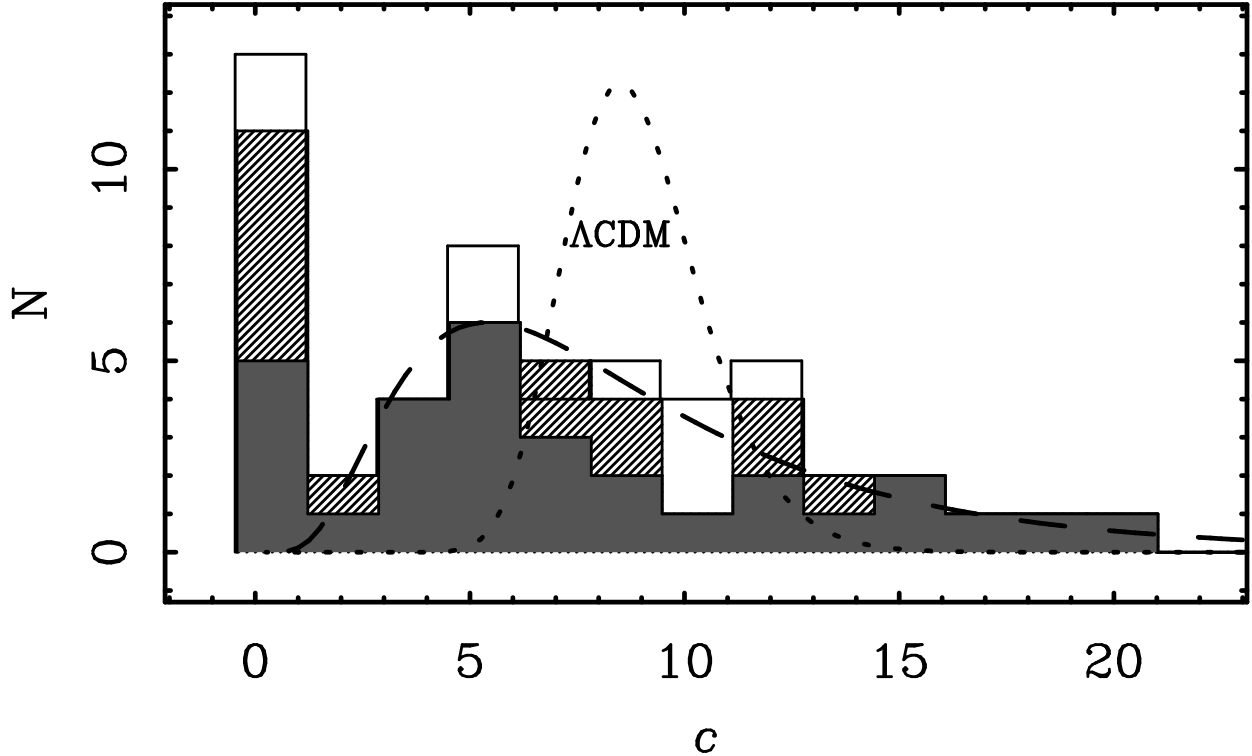}
\caption{
Histogram of the observed concentrations from the data of BMR and BB for the 
case of minimum disk.  Several histograms are shown for different
subsets of the data.  The open histogram shows all data from BMR and BB.  
The hatched histogram includes only those objects for which the NFW fits
have $\chi^2 < 2$.  The shaded histogram includes only those with
$\chi^2 < 1$.  The shape of the observed distribution is robust to these
changes.  The characteristic concentration is low ($c \approx 6$)
and the peak of extremely low concentrations ($c < 2$) never disappears.
The data look nothing like the prediction of $\Lambda$CDM (dotted line:
Jing 2000).  A crude fit of the lognormal form advocated by Jing (2000) and 
Bullock et al.\ (2001) can be made (dashed line) if 1.\ cosmological
parameters are adjusted in accordance with the observed low median
concentration, 2.\ the amount of scatter is allowed to be much larger
than found in simulations, and 3.\ the galaxies with $c < 2$ are ignored.
\label{f3}}
\end{figure}

\placefigure{f3}

The observed distribution of concentrations is robust.
Fig.\ 3 shows three different selections of the data: 1.\ all of the data
of BMR and BB (54 galaxies), 2.\ only those galaxies (45) with tolerable NFW
fits ($\chi^2 < 2$), and 3.\  only those (31) with good fits ($\chi^2 < 1$). 
While there are of course
some detailed differences between these, the overall shape of the distribution
is unaffected.  The central peak is unmoved; its width varies little.
The large peak of galaxies with very low concentrations ($c< 2$) never
disappears.  One might have hoped that by including only the best fits, a
sensible distribution would emerge, and the galaxies with unacceptably
low concentrations would disappear.  Unfortunately, simply having
a good NFW fit is no guarantee that the halo parameters correspond
to any sensible cosmology:  no cosmology ever produces halos with $c < 2$.

The failure of the data in Fig.~3 to conform to the predictions of
\LCDM\ is not a data quality issue (McGaugh \etal\ 2001; BMR; BB).
Many of the galaxies with $\chi^2 < 1$ have $\chi^2 \ll 1$,
implying that the error bars have, if anything, been overestimated. 
This makes the many objects with $\chi^2 > 1$ more problematic. 
These are generally the better data.  NFW fits with 
$\chi^2 < 1$ are usually obtained when the
uncertainties\footnote{$\Delta \chi^2$ generally favors soft cores (BMR; BB).}
are large, not because the data look like cuspy halos.
In addition to making cuts by $\chi^2$, we have also made cuts by
inclination and by subjective judgement of the data quality 
(de Blok, Bosma, \& McGaugh 2003).  The results are indistinguishable
from those presented here.

There are many galaxies with concentrations which are either too small
or too large for \LCDM.
There may be astrophysical explanations for these outlyers.
An obvious explanation for the galaxies whose concentrations are too high
for \LCDM\ would be that they are not entirely dark matter dominated.
An appropriately massive stellar disk, when subtracted off, might leave
the right amount of dark matter.  One has to be careful with such a procedure,
as a heavy disk implies some compression of the original (cosmological) halo,
which might end up with too low a concentration if one merely aims to get
the ``right'' concentration currently.  Moreover, it might be possible
to choose a stellar mass-to-light ratio
which simply gives a desired result.  It is for this reason that a large
sample of dark matter dominated galaxies is best for this experiment.

That said, it is not obvious that plausible stellar mass-to-light
ratios will reduce the implied concentrations of the high-$c$ outlyers.
In the cases where adequate photometry is available, BMR and BB present
mass models with $M_*/L_R = 1.4\; M_{\odot}/L_{\odot}$.  In none of the cases
of high concentration does the inclusion of stars of this mass-to-light
ratio significantly reduce $c$.  Implausibly large mass-to-light ratios
would be required to have an impact.  The excess number of high-$c$ galaxies
may therefore pose a genuine problem.

A more severe (and certainly genuine) problem is posed by the low-$c$ outlyers.
This can not be solved by allowing stars to have
mass, as this makes matters worse.  Indeed, contrary to the case of
high-$c$ galaxies,  attributing even a small mass to the stars can
have a significant downwards impact on $c$ if it is already low in the
minimum disk case.  More generally, these extremely low concentration
galaxies embody the soft core problem for which many explanations have
been hypothesized.  We therefore postpone discussion of this problem
until \S 4.4.

\subsection{A Limit on Cosmology}

In this section we derive a limit on cosmology by requiring that the
peak of the distribution of observed concentrations $c_p$ follow from
cosmology in the manner prescribed by NFW.
Extracting the optimal value of $c_p$ from the data is at once both
straightforward and challenging.  Straightforward, because the peak
position in Fig.~3 is well defined, and challenging, because
the data do not agree with the NFW form.  Properly, we would use the
$\chi^2(c,V_{200})$ of the fits from BMR and BB to compute the
likelihood distribution and hence the optimal $c$.  
In practice, there is essentially zero likelihood because the NFW form
provides such a poor description of so much of the data.
In order to do this exercise at all, 
we overlook this small failing of the NFW model and work directly with
the raw histogram of concentrations (Fig.~3).  This is a generous thing to do,
as it gives credit to fits which do not really fit, and the net result
remains robust because the location of $c_p$ is the same regardless
of how the data are subdivided.

In Table~1 we give several estimations of $c_p$.  These include eyeball
fits of the lognormal distribution, and the robust statistical estimators
the median and the biweight location.  
Formal fits with the lognormal function are ill-defined for the same
reason that the likelihood can not be used directly.  Nonetheless,
the eyeball fits do nicely describe much of the data:  allowing
the lognormal form to have a large scatter\footnote{The scatter
is much larger than anticipated.  A large scatter in $c$ 
should have a severe impact on the scatter in the Tully-Fisher relation. 
That it apparently does not leads to fine-tuning problems 
(McGaugh \& de Blok 1998a; Bullock \etal\ 2001).}
($\sigma_c \approx 0.6$) provides a plausible
explanation for the high-$c$ outlyers (while offering no explanation for the
low-$c$ galaxies).  The skew of the distribution tends to
make $c_p$ from the lognormal fits a bit lower than the median and
biweight location. The latter are useful statistics because they are robust
against outlyers, though even their interpretation is open to question since
the many outlyers may simply be another indication that the underlying model
is inappropriate.

\placetable{t1}

We can now place a limit on cosmology by equating $c_p$ derived from the
observations with $c_{\Lambda,12}$ from equation~(4).  In doing so, we
are making the approximation that the observed concentrations apply to
$10^{12}\;M_{\odot}$ halos rather than the particular mass scale appropriate to
each individual fit.  In practice this is a very good approximation because
the predicted $c$-$V_{200}$ relation is very flat (NFW), so the correction to
$c$ when projected to the $V_{200}$ appropriate for a given galaxy
is usually smaller than the uncertainty in $c$.
Moreover, most LSB galaxy halos should be less massive than
$10^{12}\;M_{\odot}$,
so $c_p$ should, if anything, be compared to the higher concentrations
predicted for lower mass halos.

Evaluating equation~(4) with the lognormal $c_p = 5.4$ yields
$\sigma_8 \Gamma_{0.6} = 0.15$.  For comparison, standard \LCDM,
with $\Om = 0.33$, $h = 0.66$, $n =1.03$ (Netterfield \etal\ 2001),
$\Omega_b h^2 = 0.02$ (O'Meara \etal\ 2001),
and $\sigma_8 = 0.96$ (Pierpaoli, Scott, \& White 2001) gives\footnote{
For these parameters, the ordinary shape parameter is $\Gamma = 0.18$,
slightly less than used by Pierpaoli et al.\ (2001): $\Gamma = 0.23$.}
$\sigma_8 \Gamma_{0.6} = 0.28$.  This is nearly twice the value derived here,
and predicts substantially higher concentrations: $c_{\Lambda,12} = 8.5$
(Fig.~3).  Such concentrated halos have a very distinctive dynamical
signature and would be {\it easily\/} recognized by current observations
(de Blok et al.\ 2003).

If dark matter halos were well described by cuspy halos, the peak of
the distribution of concentration parameters would provide a useful
measurement of $\sigma_8 \Gamma_{0.6}$.  However, the NFW form, upon which this
analysis is based, does not provide a good description of the data (BMR; BB).
Instead, halos with lower density cores are preferred.  Even persisting
with NFW halos as we have done here, the observations demand rather low
concentrations.  Not only is $c_p$ lower than expected, but there is also
the substantial population of galaxies with $c < 2$ which have no explanation
in the cuspy halo picture.  The data are telling us that dark matter halos
can not be as concentrated as nominally expected.  Rather than attempt to
use this method to measure cosmic parameters,
we ask: what cosmology could produce tolerable concentrations?
This leads to an upper limit on $\sigma_8 \Gamma_{0.6}$
which is quite firm as long as the NFW picture of halo formation holds.

In order to limit cosmology to parameters which might produce suitably low
concentrations, we note that 95\% of our realizations have
$c/c_{\Lambda,12} < 1.15$.  A given $c_p$ cannot originate from a cosmology
which produces halos more concentrated than this.  The largest estimate of
the peak concentration from Table~1 is $c_p = 6.4$.  Multiplying this by 1.15
to allow for the residual scatter about equation~(4)  leads to the limit
\begin{equation}
c_p < 7.4.
\end{equation}
The corresponding limit on cosmology is
\begin{equation}
\sigma_8 \Gamma_{0.6} < 0.23.
\end{equation}
This is a very hard limit, as we have adopted the largest estimate
of the peak location from Table 1 and hedged upwards from there.
The data certainly do not indicate such a large $c_p$,
and the true vale of $\sigma_8 \Gamma_{0.6}$ must be much less
if there has been no radical redistribution of mass subsequent to
halo formation.

\section{Discussion}

In this section, we discuss possible interpretations of the
concentration limit on cosmology.  These come in several basic flavors:
\begin{enumerate}
\item CDM halos must have cusps, so the stated limits hold and provide
new constraints on cosmological parameters.
\item Something (e.g., feedback; modification of the nature of dark matter)
eliminates cusps and thus the constraints on cosmology.
\item The picture of halo formation suggested by CDM simulations is wrong.
\end{enumerate}
The bulk of the dynamical data may well prefer the last of these
interpretations, potentially with drastic consequences for CDM
(McGaugh \& de Blok 1998a,b; de Blok \& McGaugh 1998;
Sanders \& Verheijen 1998; Sanders 2000; Sanders \& McGaugh 2002).
We focus the discussion here on items (1) and (2) which
have a variety of sub-flavors.

\subsection{Assumptions}

As in any study of cosmological parameters, a number of assumptions
must be made.  Before trying to sort out possible interpretations, it is
worth examining the validity of the assumptions underlying our analysis.
Everything we do here is confined to the context of the
currently standard $\Lambda$CDM cosmology.  In order to map the
rotation curve results to cosmological parameters, we have made
a number of operative assumptions:
\begin{enumerate}
\item The NFW halo is the correct prediction of CDM.
\item The concentration parameter maps to cosmology via the NFW prescription.
\item NFW halos provide an adequate description of the data.
\item LSB galaxies reside in typical halos.
\end{enumerate}

Assumption (1) has already been discussed in \S 3.1, and appears certainly
to be true modulo only the remaining debate over the precise
slope of the inner cusp.  Since $\gamma = 1$ is at the lower limit
of predicted cusp slopes, this assumption is both valid and conservative.
Steeper cusps would result in more stringent limits on cosmology.  If we accept
(1), (2) follows.

Assumption (3), that NFW halos provide an adequate description of the
rotation curve data, is the most dubious (\S 2). The most obvious interpretation
of these data is that dark matter halos do not have cusps (de Blok \etal\ 2001),
or do not exist at all (Sanders \& McGaugh 2002). 
However, the point of this paper is to explore the consequences if we
insist on retaining cuspy halos.

Assumption (4) is important because we treat the concentrations
determined from LSB galaxy rotation curves as a measure of that
produced by cosmology.  If for some reason halo concentration is
correlated with surface brightness, then the cosmological measure will be
biased.  There is little reason to suspect such a bias in theory,
and none empirically.

In most modern theories of galaxy formation
(e.g., Dalcanton, Spergel, \& Summers 1997; Mo, Mao, \& White 1998;
McGaugh \& de Blok 1998a; van den Bosch \& Dalcanton 2000)
it is the spin of the halo and not its concentration which
dictates the surface brightness.  Concentration may be correlated with
formation epoch, and scatter in the latter may well be the dominant cause
of scatter in the former (Wechsler et al.\ 2002).  It is not unreasonable
to suppose that LSB galaxies form late (McGaugh \& Bothun 1994), but this
does not happen in theories where spin is the dominant factor determining
surface brightness.  Moreover, the range of formation times discussed by
Wechsler et al.\ (2002) is far too small to explain the range of observed
concentrations.

One could of course construct a theory which
imposes a correlation between concentration and surface brightness.
The motivation to do this is not independent of the problems
discussed here, and faces a host of other problems (discussed in detail
by McGaugh \& de Blok 1998a).  Most importantly, such an approach
runs contrary to the remarkable empirical normalcy of LSB galaxies.

In terms of their physical properties (metallicity, gas content, etc.),
LSB galaxies are quite normal for their part of the luminosity function.
They adhere to the same Tully-Fisher relation as do brighter
galaxies (Sprayberry et al.\ 1995; Zwaan et al.\ 1995; Tully \& Verheijen
1997), and to the same baryonic Tully-Fisher relation
(McGaugh et al.\ 2000; Bell \& de Jong 2001; Verheijen 2001). 
This is usually interpreted to mean that LSB galaxies inhabit halos
which are similar to those of HSB galaxies of the same mass.
Invoking a correlation between concentration and surface brightness
predicts a shift in the Tully-Fisher relation between HSB and LSB galaxies
which is not observed.  

The empirical normalcy of LSB galaxies includes not just the normalization
of their asymptotic rotation velocities (the Tully-Fisher relation),
but extends also to the {\it shapes\/} of their rotation curves.   These
are quite predictable given knowledge of their luminosity distribution
(e.g., Persic \& Salucci 1991; de Blok \& McGaugh 1998;
Sanders \& McGaugh 2002).  This obedience\footnote{The adherence of LSB
galaxies to the scaling relation for rotation curve shape is also another
indication that the data do not suffer from systematic errors.}
to scaling relations established for HSB galaxies is a strong indication
that LSB galaxies are dynamically normal.
There is, therefore, no reason to suspect that assumption (4) is invalid.

\subsection{How Firm a Limit?}

Proceeding with the above assumptions, we next examine the firmness
of the concentration limit on cosmology.  So long as we insist on
having cuspy halos, this limit is valid.  Indeed, we have been quite
conservative in placing it.  On every occasion where there has been
any room to hedge, we have done so in the direction which maximized
the allowed concentration.  Hedges which act in this way include:
\begin{enumerate}
\item The adoption of minimal disks.
\item Ignoring adiabatic contraction.
\item Ignoring the angular momentum catastrophe.
\item The use of heavy halos to predict $c_{\Lambda,12}$.
\item Ignoring the low-$c$ spike in the distribution.
\item Adopting the largest estimate of $c_p$.
\item Making a generous allowance for scatter in
	the $c_{\Lambda,12}$-$\sigma_8 \Gamma_{0.6}$ relation.
\end{enumerate}
These combine to make the limit $\sigma_8 \Gamma_{0.6} < 0.23$ both
conservative and hard.  Taking the best-guess value of any of these
effects would lead to a lower value, in some cases by a large factor.

Item (1) is very generous.  Stars do have mass, but we have
pretended they do not.  Consequently, the concentrations we use do not
refer to the primordial dark matter halo as they should, but rather to
the present dark matter halo plus the stars.  Subtracting off the stars
lowers $c$, even in LSB galaxies where dark matter domination minimizes
this effect but does not entirely eliminate it.  

Items (2) and (3) depend on the galaxy formation process.  The natural
expectation is that whatever baryons collapse to form the disk will drag
along some of the dark matter (2).  This will make the present-day halo
more concentrated than the primordial halo which CHARDEN computes.  This
effect is probably modest in LSB galaxies, but should act some --- in the
wrong direction.  Item (3) is hard to quantify (Steinmetz \& Navarro 2002),
but could be quite severe, and again acts in the wrong direction.

The remaining items on the above list have been discussed as they arose
in the analysis, and we will not repeat this here.  The point is that
there are many turns where we have adopted concentrations which are,
if anything, too high.  The fact that fiducial \LCDM\ cosmologies predict
still higher concentrations emphasizes the severity of this problem.

\subsection{Implications for Cosmology}

Having reviewed our assumptions and the validity of the concentration
limit, we turn now to the first flavor of possible interpretations.
If we insist that galaxy halos must have cusps,
the limit imposed on cosmology is unavoidable.
Here we consider the implications of this limit in the context of other
cosmological constraints.  In a subsequent section we will address
the possibility of dodging these constraints by invoking processes
which might alter cusps (e.g., feedback; warm or self-interacting dark matter).

The concentration limit $\sigma_8 \Gamma_{0.6} < 0.23$
excludes a significant fraction cosmological parameter space,
including that occupied by our fiducial \LCDM\ parameters (\S 3.4).
In order to satisfy this limit, we need to reduce the
density of the universe at the time of halo collapse.  This can be
accomplished by lowering the matter density directly, or by delaying
halo formation by suppressing the power spectrum on galaxy scales.
We examine these possibilities in turn.

\subsubsection{Matter Density}

For a standard power spectrum and baryon density,
the concentration limit excludes $\Omega_m^{0.6} h > 0.28$ (Fig.~4).
This limit is fairly
restrictive, excluding some parameter combinations which are otherwise
viable.  $\Omega_m = 1$ is right out, and even $\Omega_m \approx 0.3$
cannot be sustained.  Cuspy halos and ``standard'' \LCDM\ 
are mutually exclusive. 

\begin{figure}
%\plotone{omh.ps}
\plotone{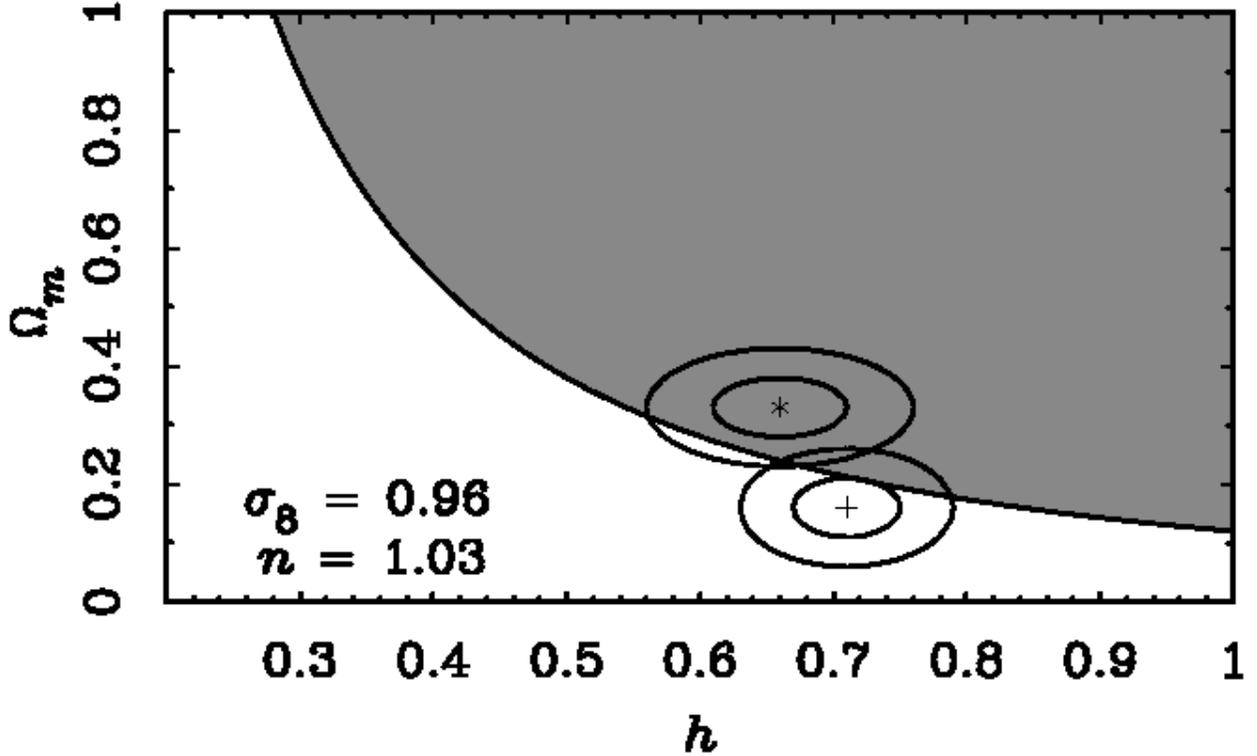}
\caption{
The region of the $\Omega_m$-$h$ plane excluded by
the concentration limit on $\sigma_8 \Gamma_{0.6}$ assuming
$\Omega_b h^2 = 0.02$. 
The entire region $\Omega_m^{0.6} h > 0.28$
is excluded if we require NFW halos to fit observed rotation
curves in the standard $\Lambda$CDM cosmology.
For comparison, we also show $1\sigma$ and $2\sigma$ error ellipses
around independent determinations of $\Omega_m$ and $h$. 
The asterisk at the point (0.66,0.33) illustrates the standard $\Lambda$CDM
value as fit by Netterfield et al.\ (2001).
Such a ``high'' density universe and cuspy halos are mutually exclusive.
the cross at the point (0.71,0.16) takes for the
Hubble constant the results of the HST key project on the
extragalactic distance scale (Sakai et al.\ 2000).
For the mass density, we take the value estimated by 
by Bahcall et al.\ (2000): $\Omega_m = 0.16 \pm 0.05$. 
It is possible to consider NFW halos for many galaxies
in such a very low density universe, though a plausible explanation
for those with $c<2$ remains wanting.
\label{f4}}
\end{figure}

\placefigure{f4}

It may be possible to salvage the \LCDM\ picture with cuspy halos if the
density is low.  For $h = 0.7$, the concentration limit requires $\Om < 0.22$.
While this is lower than usually quoted for \LCDM,
it is actually in keeping with a number of recent determinations.
Bahcall et al.\ (2000) give $\Omega_m = 0.16 \pm 0.05$. 
A similar number is also given
by Rines \etal\ (2001) --- $\Omega_m = 0.17 \pm 0.05$, and 
by Hoekstra et al.\ (2001) --- $\Omega_m = 0.13 \pm 0.07$ for a flat universe.
These independent determinations are consistent with the findings presented
here, and may be an indication of a rather small density parameter which
is acceptable to many independent data sets (Peebles 1999). 
However, it may be hard to simultaneously reconcile such a low matter density
with the requirement for both flatness $\Om + \OL \approx 1$
(de Bernardis et al.\ 2000) and the limit on the cosmological
constant from gravitational lensing, $\OL < 0.7$ (Kochanek 1996;
Cooray, Quashnock, \& Miller 1999). 

\subsubsection{Power Spectrum}

If a density parameter as low as $\Om < 0.22$ is not acceptable, the
requirement stipulated by the concentration limit may be satisfied
by a decrease in the
amplitude of the power spectrum on galaxy scales.  This can be
achieved either by a tilt or a decrease in the normalization
\norm, or some combination of both.  For example, for $\Om = 0.33$
and $h = 0.66$, we need $\sigma_8 \lesssim 0.8$ (Fig.~5).  
Lately there have been contradictory tugs on the value of the normalization,
with a combined analysis of the 2dF and CMB data suggesting
$\sigma_8 \approx 0.73$ (Lahav \etal\ 2002), while high multipole
CBI data suggest $\sigma_8 \sim 1$ (Bond \etal\ 2002).

\begin{figure}
%\plotone{s8n.ps}
\plotone{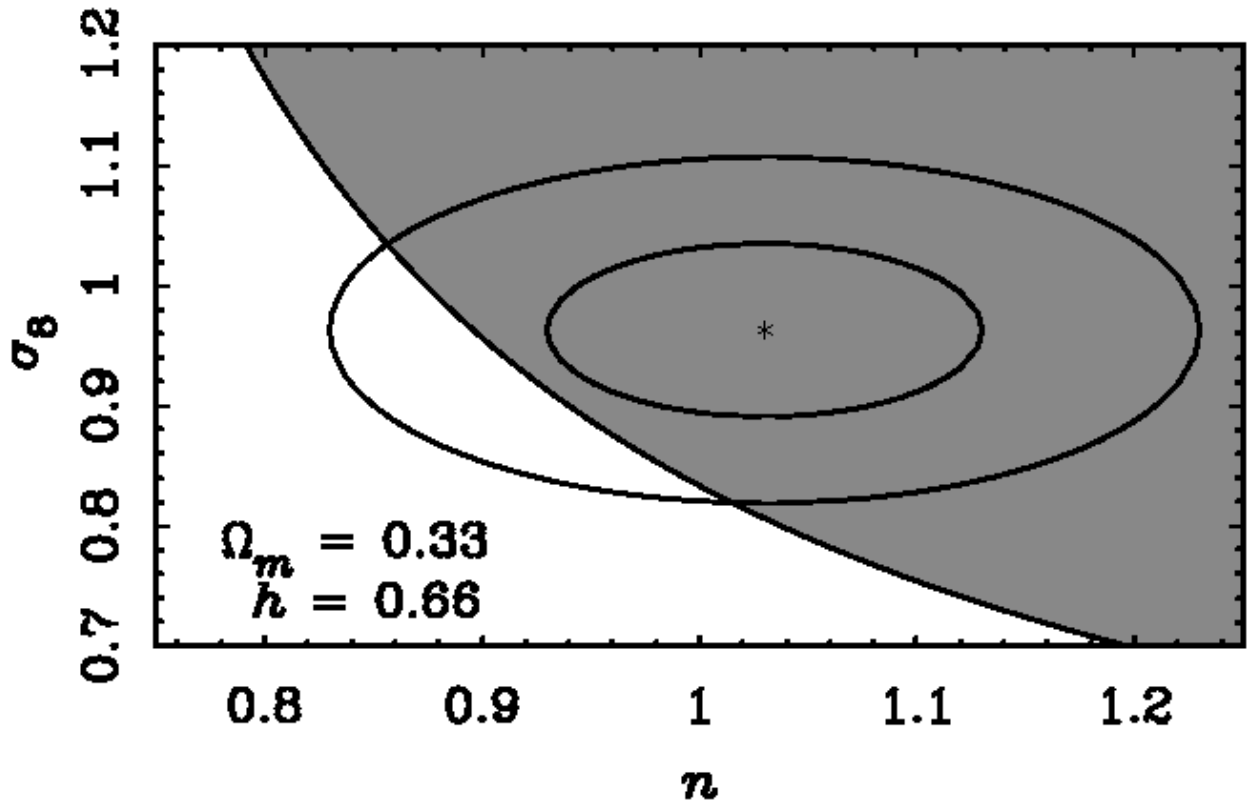}
\caption{Power spectra parameters
excluded by the concentration limit for nominal ($\Omega_m,h$) of
(0.33,0.66).  Also shown (asterisk with $1 \sigma$ and $2 \sigma$ error
ellipses) are the best estimates of these parameters from
Netterfield et al.\ (2001: $n = 1.03 \pm 0.10$) and Pierpaoli et al.\
(2001: $\sigma_8 = 0.96 \pm 0.07$).  A substantial reduction in the
amplitude of the power spectrum on galaxy scales can decrease the
concentration of dark matter halos to tolerable levels, though it leaves
open the question of why their preferred shape is not cuspy.
\label{f5}}
\end{figure}

\placefigure{f5}

The same effect could also be achieved by tilting the power spectrum
($n \lesssim 0.9$) or by introducing a break in the power spectrum at
some appropriate scale.  Such behavior is contrary to the scale-free nature
of CDM, but might occur with warm dark matter or an admixture of hot dark
matter.  This may not be necessary as a purely cosmological solution appears
to be least marginally viable.  A universe with a low mass density or
suppressed power spectrum (or both) could satisfy the concentration limit,
though it would leave open many related questions.

\subsection{Halo Modification}

This section examines the second flavor interpretation,
that some mechanism alters the initially cuspy form of dark matter halos.
This approach has the advantage that it might explain why galaxy
observations prefer halos with soft cores to those with cusps,
with little or no consequence for cosmology.  It has the rather substantial
disadvantage that some drastic and poorly understood effect must be invoked
to alter halos, thus destroying the elegance and predictive power of the
NFW paradigm.

\subsubsection{Feedback}

The mechanism most commonly invoked in this context is feedback.
This is the notion that the action of star formation, most particularly
winds and supernovae from massive stars, injects sufficient mechanical energy
into the interstellar medium to alter the surroundings.  The hope is that such
feedback might mediate between the cosmological initial state of halos
predicted by simulations and their current observed state.

The term feedback has come to be used to mean a great variety of
things.  We must immediately make a distinction between the relatively mild 
sort of feedback activity which is actually observed in galaxies, and the
explosive feedback required to address the problems posed by
soft core halos.  There is no doubt that the former does occur
(e.g., Martin 1998, 1999; Rupke, Veilleux, \& Sanders 2002),
and probably plays an important role in enriching the intergalactic medium.
However, the amount of gas involved is generally a small percentage of a
galaxy's interstellar medium, which is a small fraction of its baryonic mass,
which is a fraction of its dark mass.  The observed examples
of feedback are far too feeble to have any significant impact on the
distribution of the dominant dark mass.

In the context of galaxy formation, feedback is invoked to do a variety
of things which bear little relation to observed feedback.  Feedback from
some star formation might suppress further star formation.  If this varies
systematically with galaxy mass, it might translate the steep halo mass
function predicted by CDM into the flat observed galaxy luminosity function.
The dramatic loss of angular momentum experienced by baryons in live-halo
simulations (the ``angular momentum catastrophe'') leads to the formation
of disks which are much too small.  Feedback is invoked as a possible cure
for this, though sensible numerical implementations have no such effect
(e.g., Navarro \& Steimentz 2000).  Most importantly for our purposes here,
explosive feedback is invoked to drive out so much gas that it gravitationally
drags some of the dark mass with it, perhaps creating a soft core where
initially there had been a cusp (e.g., Navarro, Eke, \& Frenk 1996).

Supposing, for the moment, that explosive feedback might be able to turn cusps
into cores, there are two possible interpretations.  One is that {\it all\/}
galaxies are affected so that the current dark matter distribution
bears no resemblance to the cosmological prediction.  In this case, 
observations of LSB galaxies are fossil records of the mass redistribution
process with no implication for cosmology.  In effect, this invokes a
{\it deus ex machina\/} to render irrelevant all rotation curve data.
A second, less extreme possibility is that cusp destroying
feedback occurs only in {\it some\/} galaxies. 
In this case, feedback need only be invoked to explain
galaxies with $c < 2$.  The cuspy halo fits to the
remaining galaxies hold --- as do the cosmological limits derived from
them.  Since all of the estimators in Table~1 already ignore the low
concentration galaxies, the constraint $\sigma_8 \Gamma_{0.6} < 0.23$
is unaltered.

A separate question is whether explosive feedback really happens and
can have the desired effect
of converting a dark matter halo with a cusp into one with a core.
McGaugh \& de Blok (1998a) raised an empirical objection to this scheme.
The LSB galaxies in which this is an issue are quite gas rich 
(McGaugh \& de Blok 1997; Schombert, McGaugh, \& Eder 2001) --- a curious
state for galaxies which were supposed to have exploded so energetically that
so much gas was swept out that the dark matter was pulled along with it.  One
does require the expulsion of a huge amount of mass for this mechanism
to have any hope of working, as the dark matter can be dragged along with
the gas only by the weak shackles of gravity.  In order to arrive at the
current observed state, LSB galaxies must undergo a double-whammy formation
scenario.  First, an intense knot of star formation must occur inside the
primordial cuspy halo.  Feedback from this star formation must completely
detonate the baryonic component of the initial galaxy, sweeping out all
gas and hopefully converting the dark matter cusp into a constant density
core.  Subsequent to this mass redistribution, some gas must reaccrete to
reform the galaxy into its more tenuous present state.  This seems like
a lot to ask, and has the undesirable consequence of inserting an untestable
intermediary step between prediction and reality. 

The fundamental problem with invoking explosive feedback to redistribute mass
is that it is a case of the tail wagging the dog.  A small fraction of the
minority baryons --- those which form the first stars --- must have a
tremendous effect
on the majority dark matter.  This huge effect must be most severe where the
dark mass is most strongly concentrated, and mediated only by the weak force
of gravity.  

Basic physics considerations make the required mass redistribution highly
unlikely.  Numerical simulations of feedback suggest much weaker effects.
For example, MacLow \& Ferrara (2000) find that feedback can be effective
only in galaxies several orders of magnitude less massive than those considered
here, and only in ejecting gas, not redistributing dark mass. Gnedin \&
Zhao (2002) put a strict limit on the possible effects of explosive
feedback by examining the consequences of the instantaneous removal of
all gas.  They find that even this extreme fails to destroy the initial
cusp.  Hence, it seems unlikely that explosive feedback can have the mass
redistributing effects which are required to address the cusp-core problem.

\subsubsection{Other Mass Redistribution Mechanisms}

There could be mechanisms to redistribute mass besides feedback.
Weinberg \& Katz (2001) suggested that bars in disks could impart enough
angular momentum to halo particles to alter the halo mass distribution.
However, this process can only be effective when the disk is a significant
fraction of the total mass.  Hence this mechanism might at best work in
HSB galaxies like the Milky Way.  It can not, as Weinberg \& Katz note, explain 
dark matter dominated LSB galaxies unless explosive feedback is invoked
first.  Whether this mechanism is viable even in principle has been questioned
by Sellwood (2003).

Another dynamical mechanism invokes the inspiraling of supermassive
black hole pairs (Milosavljevi{\' c} et al.\ 2002).  This process might
displace up to ten times the black hole mass, and could well be important
in elliptical galaxies.  However, the black hole mass--velocity dispersion
relation (Merritt \& Ferrarese 2001) predicts very small black hole
masses for bulgeless, dynamically cold LSB galaxies:  far too small
to cause any significant redistribution of dark mass.  For example,
a galaxy comparable in mass to many of the LSB galaxies discussed here
is M33.  The limit on the central black hole mass in M33 is
$< 3000\; M_{\odot}$ (Merritt, Ferrarese, \& Joseph 2001).
Unless LSB galaxies host abnormally massive black holes,
and acquired them in pairs through mergers (which these galaxies
appear not to have experienced), this mechanism can not apply to them.

\subsubsection{Dark Matter Physics}

Mass redistribution mechanisms which invoke the interaction of baryons
and CDM appear either not to be viable, or not to apply to LSB galaxies. 
One might next consider modifying the nature of dark matter in order to
alter the cuspy halo prediction.  Such ideas fall into two broad categories:
those which prevent the formation of cusps in the first place, and those which
might reduce their concentration to more tolerable levels.  Either would
relieve or eliminate the cosmological constraints imposed here in the
strict context of pure CDM.

Perhaps the first case to consider is a mixed hot plus cold dark matter
cosmogony.  Neutrinos do appear to have mass, and so will affect structure
formation at some level.  Massive neutrinos have the effect of suppressing
the power spectrum on small scales relative to what it would have been in
their absence.  This operates in the desired direction, though a fairly
hefty neutrino fraction ($\Omega_{\nu}/\Omega_m \gtrsim 10\%$)
is probably needed to have a
significant impact.  This would require neutrinos near the current upper
bound on their mass (a few eV).  This may well be possible, but can at
best reduce the concentrations of halos a bit.  The halos should still have
cusps, not cores.  So while it may be possible in this fashion to relax
somewhat the constraints on cosmology, the more basic question about the
shape of the density distribution of the dark matter halo remains unaddressed.

An effect similar to a mixture of hot and cold dark matter can be obtained with
warm dark matter (WDM).  In this case, the mass of the particle is fine-tuned 
so that it is neither hot nor cold dynamically.  Again, the power spectrum
is reduced on small scales, and the halo profile may be affected as well
(Bode, Ostriker, \& Turok 2001).  The latter point is controversial,
as Knebe et al.\ (2002) find that halo cuspiness persists in WDM.
Observationally, there are already serious objections to WDM. 
Fermionic WDM should have
a characteristic phase space density which appears to be inconsistent
with galaxy and cluster data (Sellwood 2000; Marchesini et al.\ 2002).

The net effect of H+CDM and WDM models are similar for this problem.
Some suppression of the power spectrum on small scales makes the
cosmological limits more palatable.  The presence of soft cores in
at least some galaxies remains a difficult issue.

More radical suggestions about the nature of dark matter have also been
made to address the cusp-core problem.  These include annihilating and
self-interacting dark matter (SIDM: Spergel \& Steinhardt 2000).
Annihilating dark matter may form halos with a core (e.g., Craig \&
Davis 2001) but could over-suppress small scale power.  SIDM produces
a cusp for small interaction cross-sections, but can produce a core
with sufficiently large cross-sections (Dav\'e et al.\ 2001). 
The cross-section that works for galaxies does not work for clusters
(Yoshida et al.\ 2000), so one must invoke a velocity-dependent cross-section.
This seems rather contrived, and other objections have been raised: 
Kochanek \& White (2000) argue that the gravothermal catastrophe 
will cause SIDM cusps to steepen faster than they flatten.

There remains ample room to consider further modifications to the nature
of dark matter.  On the one hand, these seem like a more promising approach
to the problem of turning cusps into cores than does explosive feedback.
On the other hand, the problem is considerably more subtle than just
turning a cusp into a core.  It is well established observationally
that the distributions of dark and luminous matter are tightly coupled
(Sanders \& McGaugh 2002; see also McGaugh 2000). 
The cusp-core problem is just one manifestation
of this more fundamental issue.  None of the approaches we have reviewed
have made any attempt to explain the full richness of the observational
phenomenology.  SIDM interacts with
itself, but not with baryons.  It is hard to imagine how any explanation
of the coupling between baryons and dark matter can be achieved by 
modifications of cold dark matter which explicitly ignore the baryons.

\section{Conclusions}

We have examined the cosmological consequences of the
rotation curve data for dark matter dominated LSB galaxies in the
context of cuspy NFW halos.  If we insist that dark matter halos must
have the cusps suggested by current structure formation simulations,
an interesting limit on cosmology follows.  This concentration limit is
\begin{equation}
\sigma_8 \Gamma_{0.6} < 0.23,
\end{equation}
where $\Gamma_{0.6}$ is a suitably modified shape parameter (equation~3).
As a practical matter, this means either a low density universe
($\Om < 0.2$) or one with a depressed power spectrum on small scales
($\sigma_8 < 0.8$).

Such a universe is marginally inconsistent with the nominal
parameters of standard \LCDM.  However, it is consistent with a number
of recent determinations of the density parameter, and with most other
constraints.  It does therefore appear to be possible to tweak cosmology
in order to satisfy our concentration constraint.

If dark matter halos do not have cusps, then there is no constraint on
cosmology.  However, cuspy halos do appear to have become a fundamental
tenant of CDM structure formation.  If so, the implications for cosmology
are unavoidable.

Adjusting cosmological parameters can only
address the problem of halo {\it concentration\/}.
It does nothing to explain the {\it shape\/} of the radial mass distribution
in dark matter halos, which must be cuspy in the current paradigm.
In many objects with well-determined rotation curves, this appears not to
be the case.  Tweaking cosmology is a necessary step, but only a partial
solution to one aspect of a broader problem.

Various mechanisms have been proposed to convert cusps into constant density
cores.  These include explosive feedback and modifications of the nature
of dark matter (e.g., warm or self-interacting dark matter).  Some cusp
altering mechanism does appear to be necessary.  However, it is far from
obvious that any of the ideas which have been discussed so far are viable.
All suffer from serious problems, both empirical and theoretical.  At this
juncture, a satisfactory explanation of observed galaxy dynamics remains
beyond galaxy formation theory.

\acknowledgements We are grateful to many people for their support and
encouragement, most especially Vera Rubin, Albert Bosma, Jerry Sellwood,
and Jim Peebles.  We thank the referee and editor for a thorough review.
The work of SSM was supported in part by NSF grant AST0206078.

\clearpage

{\sl Note Added ---}  While this paper was in submission, some related
results have appeared.  Zentner \& Bullock (2002) have performed a similar
analysis, finding that a suppression of the power spectrum on small
scales helps with the concentration problem.  This appears to be in
complete accord with our results.  Swaters et al.\ (2002b) have very
recently provided a new analysis of rotation curves in addition to those
studies already cited.  This work appears consistent with previous results
insofar as $\chi^2$ prefers halos with cores to those with cusps.  There
are a number of objects in common with our sample; adding the new 
independent cases into Fig.~3 makes no difference to our result.

\clearpage

\begin{deluxetable}{lccc}
\tablecaption{Peak Location Estimates\label{t1}}
\tablewidth{0pt}
\tablehead{
\colhead{Sample} & \colhead{Median} & \colhead{Biweight} & \colhead{Lognormal} 
}
\startdata
All &	6.0 &	6.4 &	5.4 \\
$\chi^2 < 2$ &	5.7 &	6.3 &	5.4 \\
$\chi^2 < 1$ &	5.7 &	6.2 &	5.3 \\
\enddata
\end{deluxetable}

\clearpage

\clearpage

\clearpage

\clearpage

\end{document}